# HoNVis: Visualizing and Exploring Higher-Order Networks

Jun Tao* Jian Xu† Chaoli Wang‡ Nitesh V. Chawla§

University of Notre Dame

**Abstract**

Unlike the conventional first-order network (FoN), the higher-order network (HoN) provides a more accurate description of transitions by creating additional nodes to encode higher-order dependencies. However, there exists no visualization and exploration tool for the HoN. For applications such as the development of strategies to control species invasion through global shipping which is known to exhibit higher-order dependencies, the existing FoN visualization tools are limited. In this paper, we present HoNVis, a novel visual analytics framework for exploring higher-order dependencies of the global ocean shipping network. Our framework leverages coordinated multiple views to reveal the network structure at three levels of detail (i.e., the global, local, and individual port levels). Users can quickly identify ports of interest at the global level and specify a port to investigate its higher-order nodes at the individual port level. Investigating a larger-scale impact is enabled through the exploration of HoN at the local level. Using the global ocean shipping network data, we demonstrate the effectiveness of our approach with a real-world use case conducted by domain experts specializing in species invasion. Finally, we discuss the generalizability of this framework to other real-world applications such as information diffusion in social networks and epidemic spreading through air transportation.

## 1 Introduction

Modern day systems are complex, whether they are movements of hundreds of thousands of ships to form a global shipping network [14], powering the transportation and economy while inadvertently translocating invasive species; interactions of billions of people on social networks, facilitating the diffusion of information; or complex metabolic systems representing rich cellular interactions.

The complex systems are often represented as networks, where the components of the system are represented as nodes and the interactions among them are represented as edges or links. This network based representation facilitates subsequent analysis and visualization. For example, the global shipping activities are usually represented as a global shipping network, with ports as nodes, and the amount of traffic between port pairs as edge weights [16]. Traditionally, creating networks from such ship movement data has followed the port-to-port movement of a ship, and ignores the historic trajectory of the ship. This becomes extremely limiting as it has been observed that ship movements actually depend on up to *five* previously visited ports [30]; other types of interaction data from communication to transportation often exhibit *higher-order dependencies* [24, 9]. Therefore, when representing data derived from these complex systems, conventional network representations that implicitly assume the Markov property (i.e., *first-order dependency*) can quickly become limiting, undermining subsequent network analysis that relies on the network representation.

To address this problem, prior work has proposed the use of *higher-order network* (HoN) to discover higher-order dependencies and embed conditional transition probabilities into a network representation [30]. For the global shipping network example, instead of mapping every port to a single node, each higher-order node in HoN encodes not only the current step (the port that a ship currently stays) but also a sequence of previous steps (the ports that a ship visited before arriving at the current port). Therefore, the transitions among nodes in a HoN are now conditional, and are able to reproduce complex ship movement patterns more accurately from the raw data. HoN features direct compatibility with the existing suite of network analysis methods, such as random walking, clustering, and ranking, thus serving as a powerful tool for modeling the increasingly complex systems.

HoN is the correct way of representing complex systems that defy the first-order dependency assumptions. Despite the importance of HoN and its applicability to network analysis, there has not yet been a visualization tool that can handle the richness of the HoN representation. In this work, we team up with two domain experts in network science and marine ecology and develop a visual analytics framework, named HoNVis, to facilitate the exploration and understanding of HoN. The global shipping network, being an important application of HoN for the study of invasive species, is used as a case study and for illustration throughout this paper, although the general approach we take can be applied to other types of HoNs. We focus on the formation and impact of higher-order nodes, e.g., why a higher-order node exists in a HoN and how the species may propagate from a port to other ports given the previous steps? Specific to the shipping network case study, we aim to answer these questions through a three-step exploration process: 1) *global identification of ports of interest*, 2) *detailed observation of the connections of an individual port*, and 3) *tracing the propagation of invasive species from port to port through shipping*. Accordingly, we lay out the design of HoNVis in Figure 1. The input data are converted to the FoN and dependencies are extracted to construct the HoN. From these network representations, we identify nodes of interest. The visualization includes five coordinated views: *geographic view* and *table view* show information related to a single node; *dependency view*, *subgraph view*, and *aggregation view* show connections among multiple nodes. Together these five views enable users to explore higher-order nodes and their dependencies, allowing insights to be gained from this comprehensive system.

## 2 Related Work

**HoN Visualization.** HoN visualization is sporadic in the literature. Blaas et al. [6] proposed to visualize higher-order transitions by connecting nodes using higher-order curves. By following a smooth curve from one end to the other, one can identify which nodes are associated with higher-order transitions and what are the orders of the nodes. Rosvall et al. [24] grouped higher-order nodes by their current nodes and drew directed edges between connected nodes. The higher-order nodes representing the same physical locations are placed in one circle to build the correspondence. This approach, although intuitive, does not scale beyond the second or-der, nor when more than a dozen of higher-order nodes representing the same location coexist. HoN is also used as an analysis tool in unsteady flow visualization for better workload distribution. Zhang et al. [32] employed high-order dependencies to estimate the destinations of particles given their previous locations, providing more accurate information about which data blocks to load at the next step.

*e-mail: jtao1@nd.edu
†e-mail: jxu5@nd.edu
‡e-mail: chaoli.wang@nd.edu
§e-mail: nchawla@nd.edu

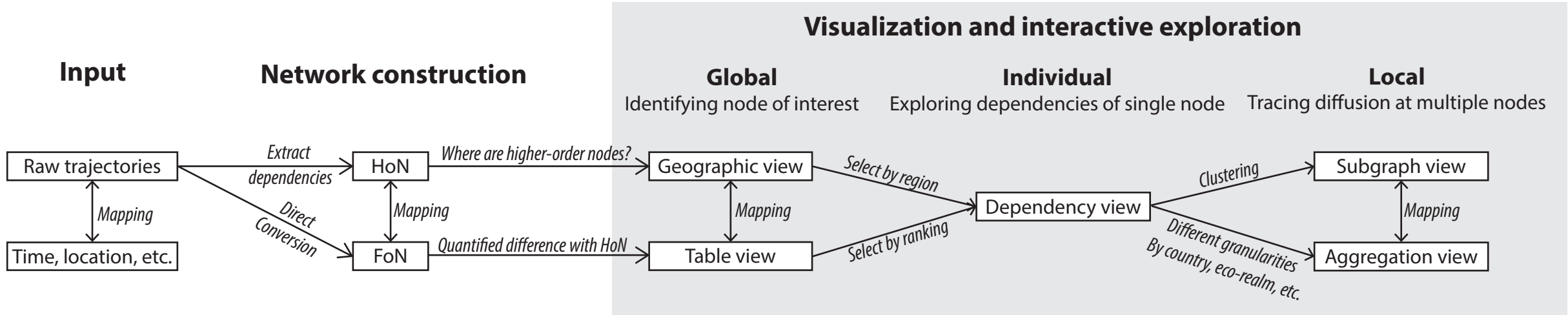

Figure 1: The framework of HoNVis design. FoN and HoN are converted and extracted from the raw trajectory data, from which we identify nodes of interest. Five linked views are designed to enable the interrogation of single and multiple nodes.

**Visual Path Analysis and Graph Comparison.** For first-order networks, quite a few works have been presented on visual path analysis and graph comparison. We refer readers to survey papers [3, 27, 2, 28, 5] for a comprehensive overview. Bodesinsky et al. [8] proposed an interface with coordinated multiple views to explore sequences of events. The event view visualizes each sequence of events as horizontally aligned bars. Event patterns are summarized in a pattern overview from which users can query a certain pattern to highlight the recurring instances in the event view. Partl et al. [22] designed Pathfinder to analyze paths in multivariate graphs. A node-link diagram visualizes the paths between queried nodes, and a ranked list shows the attributes associated with the nodes. Wongsuphasawat et al. [29] presented LifeFlow to study and compare multiple event sequences. Each sequence is displayed in a horizontal bar, and the events in different sequences are aligned vertically for easy comparison. Detailed event information for each sequence is displayed as a list.

**Movement Data Visualization.** Spatiotemporal movement data (e.g., traffic and trajectory) are often encoded as conventional graphs, where each node represents a location and an edge represents the traffic volume between two locations without distinguishing their previous locations. Guo [12] used the location-to-location graph to visualize population migration in the United States. The spatial regions are partitioned to form hierarchies and support node aggregation at the regional level. The flow is clustered based on the associated variables, such as the number of migrants for different ages and income levels. von Landesberger et al. [26] presented the MobilityGraph to visualize mass mobility. They also grouped the regions for clearer observation. To obtain a common movement pattern, a temporal clustering is performed based on the graph's feature vectors generated in different time spans. However, both works do not consider higher-order dependencies and therefore, they are not able to answer the questions such as how many migrants in Chicago who came from Los Angles would finally move to New York City.

## 3 Background

**Conventional FoN.** Usually the network representation of a complex system (e.g., global shipping) is not directly available, but needs to be constructed from the raw event sequences (e.g., ship movements) produced by the system. The conventional approach to construct a network from the raw data is to count the number of observed interactions between entities as the edge weights between node pairs. For example, given the observed ship movements shown in Figure 2 (a), a conventional shipping network is constructed as shown in Figure 2 (b), with every node representing a port and every edge representing the amount of traffic between a pair of ports. Since only direct movements are preserved in the network structure as pairwise connections, this approach implicitly assumes the Markov property, i.e., a ship's probability of moving from the current port $i_t$ to the next port $i_{t+1}$ is proportional to the edge weight $w(i_t \rightarrow i_{t+1})$. In the example of Figure 2 (b), a ship is equally likely to move to ports $X$ and $Y$ from port $M$, regardless of from where the ship coming to $M$. However, from Figure 2 (a), it is apparent that ships coming from $A$ to $M$ are more likely to go to $X$, and ships coming from $B$ to $M$ are more likely to go to $Y$. Such important information about higher-order dependencies is lost using the conventional approach.

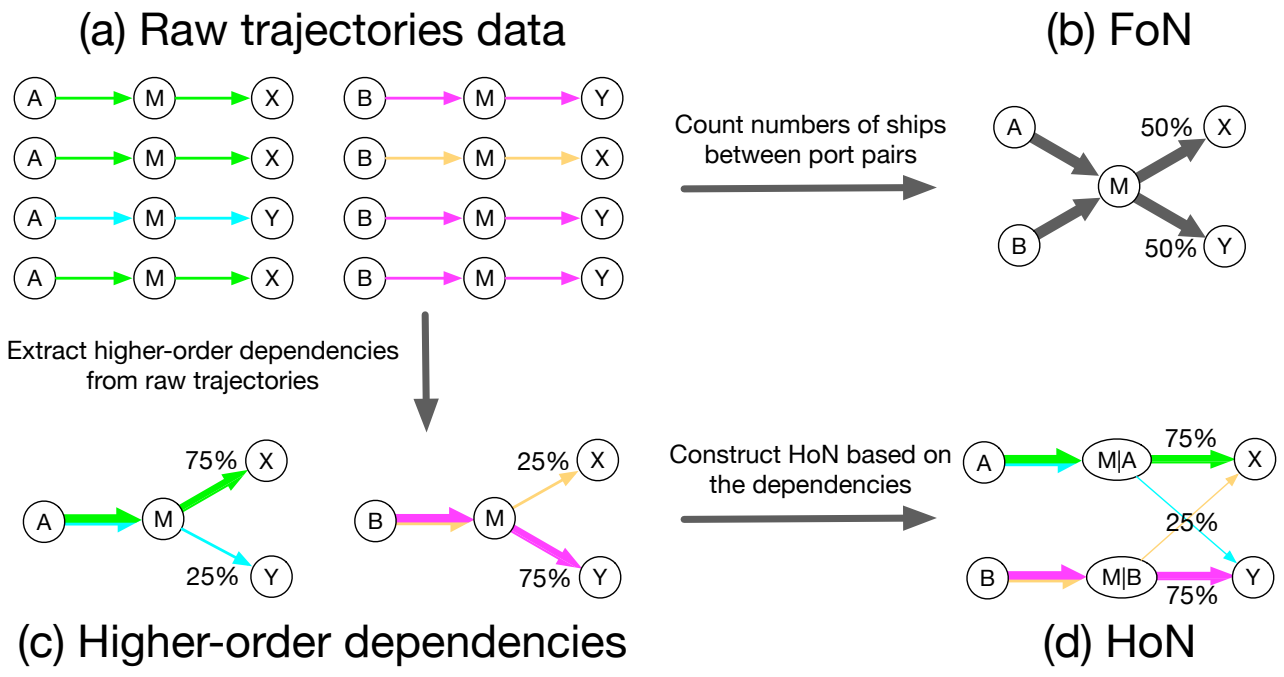

Figure 2: (a) An example of raw trajectory data. (b) Construction of the FoN from raw trajectory data. (c) Extraction of higher-order dependencies from raw trajectories. (d) Construction of the HoN from higher-order dependencies.

**Construction of HoN.** It has been shown that higher-order dependencies exist ubiquitously in flow dynamics such as ship movements, air traffic, and web clickstreams. The study on how to extract and represent arbitrary higher-order dependencies in networks, however, has just emerged. Recently, Xu et al. [30] proposed an approach that extracts higher-order dependencies from raw event sequences and embeds those dependencies into a FoN. As illustrated in Figure 2 (c), in the context of global shipping, the method first evaluates if knowing the ship came from $A$ to $M$ significantly changes the probability distribution of the ship's next step from $M$. If the change, as measured by the Kullback-Leibler divergence (KLD) [18], is significant, as the case shown in Figure 2, it suggests that the ship movements depend on not only the current port $M$ but also the previous port $A$. This comparison is iterated recursively to extract higher-order dependencies. Next, in the network representation, instead of mapping every port to a single node, every node represents the current port given a short sequence of previous ports. For example, in Figure 2 (d), the port $M$ is now broken down into two higher-order ports $M|A$ and $M|B$, such that ships coming from different ports to $M$ can have different probability distributions of choosing the next port to visit.

**How Does HoN Influence Network Analysis?** An important property of HoN is that its data structure — nodes connected by edges — is consistent with the conventional FoN (the only difference is the node labeling), making HoN directly compatible with the whole existing network analysis toolkit. When ship movements are simulated on HoN as random walking, the transition probability between ports will be

$$p(X_{t+1} = \boldsymbol{i}_{t+1} | X_t = \boldsymbol{i}_t) = \frac{w(\boldsymbol{i} \to \boldsymbol{i}_{t+1})}{\sum_{\boldsymbol{j}} w(\boldsymbol{i}_t \to \boldsymbol{j})} \quad (1)$$

where the current port $\boldsymbol{i}_t$ denoted in bold can be a higher-order node in the form of $[i_t | i_{t-1}, i_{t-2}, \dots]$, e.g., $M|A$ or $M|A,C,E$. Therefore, although Equation 1 appears to be Markovian, arbitrary orders of dependencies can be incorporated into the equation, refining the movement patterns of the simulated ship movement. This flexibility of HoN to embed *variable orders*, on the other hand, brings in new challenges to visualization as there can be tens to hundreds of nodes of variable orders representing the same physical port.

The HoN, being a more accurate representation of flow dynamics in the raw data, serves as a better foundation of subsequent network analysis. By following the auxiliary higher-order nodes and edges, random walkers on the HoN representation of global shipping demonstrate at least *twice* the accuracy on simulating the actual ship movements than on the FoN [30]. Furthermore, while the movement flows are "memoryless" and are mixed in every step on a FoN, on the HoN the flows are more clearly distinguished. That is, random walkers on the HoN have higher certainty in making every step, leading to significantly lower *entropy rates* [24, 30]

$$H(X_{t+1}|X_t) = -\sum_{\boldsymbol{i},\boldsymbol{j}} \pi(\boldsymbol{i}) p(\boldsymbol{i} \to \boldsymbol{j}) \log p(\boldsymbol{i} \to \boldsymbol{j}) \quad (2)$$

where $\pi(\boldsymbol{i})$ is the stationary distribution at node $\boldsymbol{i}$ and $p(\boldsymbol{i} \to \boldsymbol{j})$ is the transition probability from node $\boldsymbol{i}$ to node $\boldsymbol{j}$ as computed in Equation 1. The changes of random walkers' behavior on the HoN also influence the results of important network analysis methods such as PageRank [21] for ranking, which relies on random walkers to simulate movements in the network. For example, these clustering methods are based on the intuition that a random walker is more likely to move within the same cluster rather than between different clusters. In the HoN shown in Figure 2 (d), port $X$ receives more traffic from port $A$ than from port $B$, thus $X$ and $A$ are more likely to be clustered together. On the contrary, in the FoN shown in Figure 2 (b), $A$ and $B$ appear to be equivalent to $X$, regardless of the indirect flow patterns. For the study of invasive species, the clustering result on HoN provides more insight, since port $X$ is more susceptible to species originating from $A$ carried via indirect shipping.

## 4 Design Rationales

We invited two domain experts from the NSF Coastal SEES collaborative research project, who specialize in data mining and the application of marine ecology, and hold positions in R1 universities. They had spent five years on the modeling of species invasion, and had published works in interdisciplinary journals and top conferences in the domain. The experts noticed that indirect species flow through shipping exhibit higher-order dependencies [30], but had been using the FoN for visualization and control strategy development due to the lack of tools for the HoN. In this section we review the background of species invasion research, then identify requirements to guide our design of HoNVis.

### 4.1 Application Background

The ever-increasing human activities unintentionally facilitate the transportation of species, which may outcompete native species and cause substantial environmental and economical harm. The annual damage and control costs of invasive species in the United States is estimated to be more than 120 billion US dollars [23]. The global shipping network is the dominant vector for the unintentional introduction of invasive species [20]: species "hitchhike" on ships from port to port in ballast water or via hull fouling [10]. Understanding the global shipping network is crucial for devising species control management strategies. The data mining community has recently produced promising observations on the global shipping network [31]. For example, several clusters of ports which are loosely connected to each other are revealed in the global shipping network. Targeted species management strategies can be devised toward the loose connections among the clusters to prevent or slow down the species propagation from one cluster to another.

However, even the state-of-the-art research still faces unresolved challenges. For example, the recent network approach by Xu et al. [31] uses the FoN to model the species flow between port pairs; it is unclear from the FoN how species may propagate after multiple steps, and it is impossible to know which port or pathway plays an important role connecting different clusters, eco-regions, or countries. Therefore, the ability to explore the process interactively is important for the development of species management strategies.

Meanwhile, the FoN that was used to model and visualize global shipping is an *oversimplification* of higher-order dependencies that exist ubiquitously in ship movements and species flows [30]. In the iconic work of Kaluza et al. [16], a global cargo ship network was built by taking the number of trips between port pairs as edge weights, while multi-step ship movement patterns were ignored. From the visualization of the FoN thus built, one cannot tell if ships coming from Shanghai to Singapore are more likely to visit Los Angeles or Seattle. Such higher-order dependencies in networks are crucial for accurately modeling the flow of ships and species. However, no such a tool currently exists.

Finally, although it has been shown how ship types, ship sizes, geographical locations and seasonality can influence the structure of the first-order global shipping and species flow patterns [31], there has been no discussion on how such factors influence higher-order shipping patterns. It is unknown whether higher-order movement patterns are mainly formed by oil tankers, or located at estuaries, or appear mainly in winters. Such information can provide insight in revealing the driving forces behind the formation of higher-order dependencies in ship movements, and aid the development of invasive species management strategies.

### 4.2 Design Requirements

Given the gap between the demand to visualize higher-order dependencies in global shipping and the lack of HoN visualization tools, we identify key requirements for our visual analytics system.

**R1. Create a mapping between the HoN and FoN, and quantify the differences.** The experts expect to see geographical locations of ports and their connections on a map, in order to select ports at places of interest; the experts want to know if higher-order dependencies are more likely to exist in certain geographical locations (e.g., canals and straits). Additionally, the experts expect to learn how do the FoN and HoN representations compare with each other in terms of network properties such as port centralities.

While the HoN contains richer information, the FoN has the simplicity of one-to-one mapping from nodes to geographical locations on a map. To combine the advantage of both representations, we map the structure of HoN back to the FoN when visualizing it on the map, and assign scalar values to the corresponding nodes and edges in the FoN for comparison. The comparison can be defined in multiple ways depending on the exploration goal. By default, we quantify the difference of the transition probabilities between the HoN and FoN. The difference can also be quantified by comparing the network analysis results. For example, domain experts are interested in the nodes with the largest PageRank [21], which effectively simulate the flow of invasive species; the PageRank differences can help to identify ports with underestimated risks in FoN. In brief, mapping the difference or important values to FoN provides clearer observation on the map view and allows users to effectively identify and select the regions of interest for further exploration.

**R2. Provide aggregation view of the higher-order nodes.** The experts would like to explore port connections at different granularities, such as connections among countries, continents, eco-regions, eco-realms, etc. Therefore, the higher-order nodes should be aggregated and visualized for high-level knowledge discovery. For exam-ple, it should provide information such as how many nodes with the highest order exist in an eco-realm (to reveal geographical distribution of higher-order dependencies), how many pathways incorporated in higher-order nodes navigate through multiple eco-realms (to identify non-indigenous species diffusion pathways), and so on. The level of aggregation should be flexible so that users can observe the connections at different granularities, such as countries, continents, eco-regions, eco-realms, temperature and salinity ranges, etc.

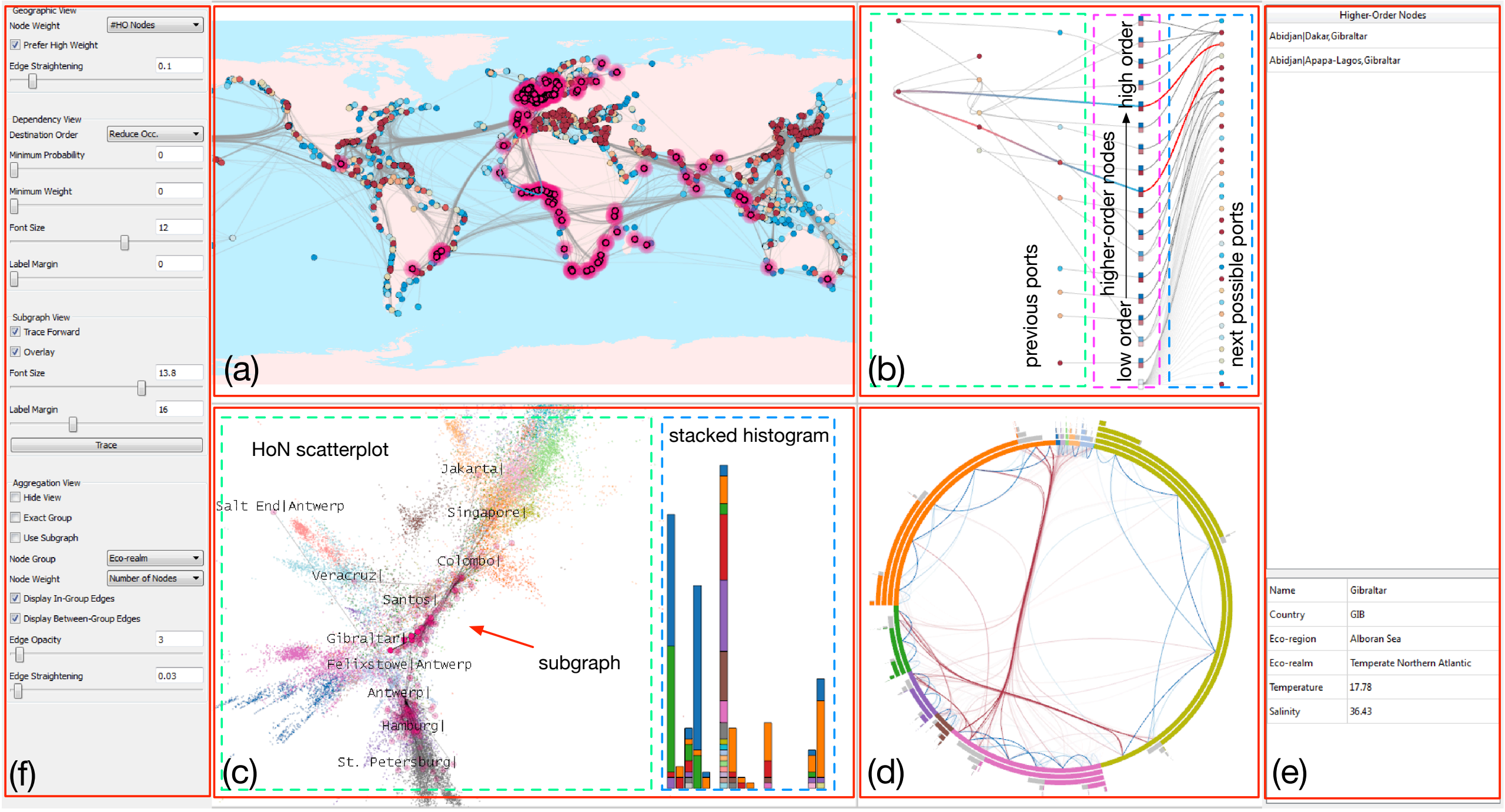

Figure 3: The overview of HoNVis, our visual analytics system for exploring the global shipping higher-order network. (a) Geographic view. (b) Dependency view. (c) Subgraph view. (d) Aggregation view. (e) Table view. (f) Parameter panel.

**R3. Visualize higher-order dependencies associated with a given port.** The experts first want to know that given a port, *how* do the previous steps change a ship's choice of the next step. For example, ships currently at Singapore may have equal probabilities of going to Los Angeles and Seattle. The experts wonder if ships coming from certain ports to Singapore will make them more likely go to Los Angeles, and how much the difference is. Meanwhile, the experts want to know if certain features correlate with the existence of higher-order dependencies. For example, are higher-order dependencies mainly associated with certain types of ships (such as oil tankers), or certain geographical locations (such as canals)?

Therefore, when a port of interest is designated, a subgraph of HoN containing all higher-order nodes and edges associated with the port of interest should be generated. The transition probabilities from different higher-order nodes to the next node should be represented, in order to show how the previous ports a ship has visited may influence the ship's next step. Additionally, the attributes of ships corresponding to the transitions should be shown, such that users may discover certain higher-order movement patterns exclusively associated with certain types of ships, particular months, and so on. For example, if the ships moving between two ports are mostly passenger ships, the ship is likely to return to the previous port, since passenger ships are likely to move between two ports instead of among multiple ports. Therefore, we should encode these attributes associated with transitions, so that once transitions of interest are identified, users can observe the corresponding attributes.

**R4. Visualize and expand a subgraph.** In the context of invasive species studies, the experts hope to see if higher-order dependencies are evenly distributed in the network or only exist in certain groups of tightly connected ports. The experts also expect to visualize and expand a subgraph of invaded ports to understand how invasive species propagate from a given port. The expansion should be performed forward or backward to cover more nodes along paths of interest. This allows interactive exploration and facilitates case studies on studying the species flow along certain shipping pathways. To understand the influence of these paths to the entire network, such as which are the important pathways that connect different clusters of ports, visual connections should be established between the subgraph and the entire network.

## 5 System Description

We design five coordinated views to meet the design requirements stated in Section 4. The five views of our HoNVis are: 1) a *geographic view* where the geographical locations of ports and the connections among them are displayed; 2) a *dependency view* that shows all the higher-order nodes associated with a given port, as well as the previously visited ports and the next possible ports to visit (Section 5.1); 3) a *subgraph view* that compares a user-generated subgraph with the graph showing the entire HoN (Section 5.2); 4) an *aggregation view* that visualizes higher-order dependencies under a certain aggregation criterion (Section 5.3); and 5) a *table view* that displays the detailed text information of a port or the current user exploration status. Users can hide the aggregation view to leave more vertical space for the dependency view. All these five views are linked together through brushing and linking. Labels of higher-order nodes/ports will be shown in the dependency view and subgraph view, since they cannot be inferred from the respective layout.

With HoNVis, a typical user workflow is as follows. Users start from the geographic view and aggregation view. In the geographic view, they identify through visual encoding (red to gray to blue for high to medium to low), the ports with more higher-order nodes or ports whose rankings change the most in the HoN compared to the FoN. In the aggregation view, both the current nodes and their previous steps are aggregated according to a given criterion. For example, when the entire network is aggregated at the eco-realm level, users can efficiently identify the higher-order nodes whose previous steps contain ports in other eco-realms, suggesting non-indigenous species introduction pathways. Users may then specify a port for individual port investigation: all higher-order nodes containing pathways leading to the given port will be visualized in the dependency view, showing how ships or species coming from different pathways to the current port will have different probability distributions of choosing the next port. Assuming a potentially invasive species in the current port, users can also trace species diffusion in the subgraph view, and understand how the species may propagate to different clusters of ports. Starting from the higher-order nodes directly related to the specified port, users can expand the subgraph of invaded ports by tracing forward or backward and including the nodes visited. This stepwise expansion gradually fills the gap between the one-step neighborhood of the selected port and the entire global shipping network, which helps users evaluate the impact of a port or a higher-order node at a larger scale. After each user operation, we use animated transition to emphasize the changes in other views, indicating where to explore in the next step. In the following, we describe the dependency view, subgraph view, and aggregation view. The other two views (refer to Figure 3 (a) and (e)) are omitted as their design and roles are straightforward.

### 5.1 Dependency View

Given a set of higher-order nodes, the dependency view shows the connections among previously visited ports and next possible ports to visit. It corresponds to the design requirements **R1** and **R3**. The higher-order nodes being investigated can be the higher-order nodes associated with a port selected in the geographic view, or multiple higher-order nodes contained in an aggregated node selected in the aggregation view. The transitions between the higher-order nodes and their next possible ports can be filtered by the probabilities or the number of ships associated with the transitions. This produces a compact visualization allowing the more important transitions to be observed clearly. A set of attributes is assigned to the ports, providing visual hints to guide the exploration. These attributes include computed ones (e.g., PageRank in the FoN, aggregated PageRank in the HoN, and the number of associated higher-order nodes) and the geographical properties (e.g., temperature, salinity, and eco-realm).

**Higher-Order Nodes.** Each higher-order node is displayed as a rectangle, as shown in Figure 3 (b). Each rectangle is divided into two boxes: the upper and lower boxes. The upper box indicates the entropy of transition probabilities starting from the higher-order node, where blue/white corresponds to low/high entropy (*low* entropy corresponds to *high* certainty). The lower box indicates the KLD of the transition probability distributions of the higher-order node and its corresponding first-order node, where red/white corresponds to high/low KLD. These two properties are of particular interest, since the first one represents the *certainty* of the next port to visit given the higher-order dependency and the second one represents the *difference* between the higher-order node and its corresponding first-order node. Therefore, distinct higher-order patterns significantly different from first-order ones show a combination of blue and red boxes and can be identified at a glance. In Figure 3 (b), we observe considerable blue/red combinations, indicating higher-order dependencies of potential interest that are not captured in the FoN. Higher-order nodes with high entropy or low KLD values, though less interesting by themselves, are indispensable for bridging the connection of other higher-order nodes.

If the number of higher-order nodes is large, we only display the lower KLD boxes of nodes, since KLD is the deciding factor for extracting higher-order dependencies and is more relevant to the formation of higher-order nodes. The higher-order nodes are lined up according to their current ports and orders: the nodes with the same current port are contiguous and the node with highest/lowest order is placed at the top/bottom of that contiguous space.

**Previous Ports.** We display the previous ports as circles to the left side of the higher-order nodes, as shown in Figure 3 (b). For each higher-order node, we draw a smooth high-order Catmull-Rom spline to connect its corresponding ports in the visit order for clear observation, as suggested by Blaas et al. [6]. The curves exhibit color transition from red to blue, indicating the visit order of ports (i.e., red indicates the port visited first and blue indicates the current port).

We determine the layout of the previous ports using a simple heuristic: their $x$-coordinates are determined by their earliest appearance in any higher-order nodes; and their $y$-coordinates are determined by the average $y$-coordinates of the higher-order nodes containing them. The ports that are placed at the same locations are moved vertically to resolve the conflict. In Figure 3 (b), we find that the ports are aligned from left to right in their visit order for most higher-order nodes. The ports associated with individual second-order nodes are mostly placed at the lower part of the dependency view and the ports associated with more higher-order nodes are mostly placed at the upper part. More sophisticated algorithms exist for drawing directed graphs, but they tend to increase the horizontal span in order to better preserve the order of nodes, which may not be ideal in our scenario given the limited screen space.

**Next Possible Ports.** We display the next possible ports as circles to the right side of the higher-order nodes, as shown in Figure 3 (b). The opacity of an edge connecting a higher-order node and a next possible port indicates the corresponding transition probability. In Figure 3 (b), since most edges associated with higher-order nodes are dark, their next steps to take are fairly certain. Furthermore, the edges associated with the first-order node at the bottom share similar light colors, which indicates that the next possible ports will be visited with similar probabilities.

The next possible ports can be lined up to reduce edge crossing or reflect a user-specified property. To reduce edge crossing, we first estimate the $y$-coordinate of a port using the average $y$-coordinates of the higher-order nodes connecting to that port weighted by their respective transition probabilities. Thus, a port will be placed closer to the higher-order nodes that are more likely to transit to it. Then, all ports are evenly spaced to span the entire screen space along a vertical line, preserving their estimated $y$-coordinates. Users can also arrange the ports according to an associated property. This facilitates the identification of transitions related to certain characteristics (e.g., high temperature or a certain eco-realm).

**Interaction.** Users can select a previous port in the dependency view for investigation. The curves associated with that port will maintain their colors, while the other curves will become gray. In the table view, we display the names of the higher-order nodes containing the selected port and the information of this port. In the subgraph view, the subgraph will be updated as well, so that users can study the propagation pattern given that port as a previous node. Users can further select a set of next possible ports. To provide detailed information, we display two histograms of ship types and temporal activities of the transitions between the selected higher-order nodes and the next possible ports.

### 5.2 Subgraph View

The subgraph view visualizes a subgraph of the HoN in the context of the entire network, corresponding to the design requirement **R2**. It shows the topological proximity of ports, and allows users to expand the subgraph of invaded ports to explore how the invasive species will propagate over the network. The entire HoN is described by a layout of the network using ForceAtlas2 [15]. Meanwhile, the structural organizations of HoN also influence the propagation dynamics. For example, the global shipping network is naturally organized into multiple communities; in each community the ports are tightly coupled by shipping traffic. Once a given species is introduced to a community, the species will propagate through the whole community shortly. Therefore, locating the *entry points* and *pathways* to communities is essential to devising species con-trol strategies. We apply the widely-used Louvain method [7] for community detection, using edge weights and the default resolution of 1.0. Note that higher-order nodes representing the same port could belong to different communities, which naturally yield overlapping clusters and indicate how certain ports may be susceptible to multiple sources of species invasion.

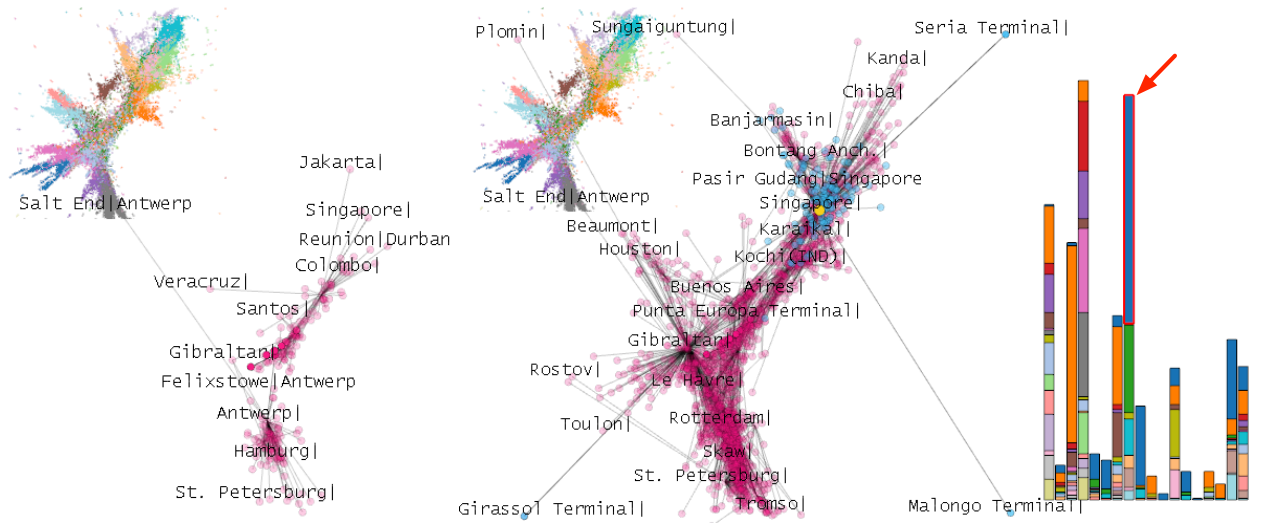

(a) (b)

Figure 4: The subgraph view. (a) HoN scatterplot and subgraph. (b) HoN scatterplot, subgraph expanded from the subgraph shown in (a), and stacked histogram showing node contribution.

We visualize the entire HoN using scatterplot, where each point represents a node in HoN, colored by the community of that node. The edges in the HoN are ignored for clutter reduction. The subgraph is then displayed on top of the scatterplot. Each node in the subgraph is drawn as a semi-transparent circle, whose center is placed at the corresponding point in the scatterplot. The transparency of a circle indicates the probability of the corresponding node being reached during the expansion of the subgraph. An edge in the subgraph is drawn as a straight line with transparency indicating the corresponding transition probability. In Figure 3 (c), the subgraph expanded from the two higher-order nodes selected in the dependency view is displayed on top of the HoN scatterplot. We can see that the subgraph mostly covers the lower right branch and the lower middle region of the network. As an option, users can choose not to overlay the subgraph and the HoN scatterplot. In that case, the HoN scatterplot will be displayed in the top-left corner of the subgraph view, as shown in Figure 4 (a). Without the overlay, the nodes in the subgraph can be observed more clearly, but the covered regions can only be roughly interpreted.

**Subgraph Expansion.** Subgraph expansion is performed by tracing from the nodes in the current subgraph and including the nodes reached during the tracing. Users can trace backward to find out through which nodes the subgraph can be reached or trace forward to explore the nodes that will be reached from the nodes in the current subgraph. The subgraph expansion procedure starts from a set of higher-order nodes selected in the other views. The initial probability of reaching a node is proportional to the number of ships leaving/arriving that node when tracing forward/backward. After each tracing step, the probability of reaching a node $n_i$ will be updated to $\sum_{n_j \in N(n_i)} p(n_j)p(e_{ji})$, where $N(n_i)$ is the set of nodes from which $n_i$ will be reached, $p(n_j)$ is the probability of $n_j$ being reached, and $p(e_{ji})$ is the transition probability from $n_j$ to $n_i$. The expansion can be observed in both the HoN scatterplot and the geographic map, where the ports associated with any node in the subgraph is highlighted. A tracing step is only performed when users click the "Trace" button in the parameter panel. This allows users to observe the propagation pattern in a stepwise manner.

**Identification of Contributing Nodes.** By *contributing nodes*, we mean the nodes that lead to the coverage of a certain community or certain regions in the HoN. The contribution of a node $n$ to a community $c$ is measured by the number of nodes in $c$ that are reached directly through $n$ for the first time. The total contribution of a node $n$ is the summation of its contributions to all communities. We choose to visualize twenty nodes with the highest total contributions using a stacked histogram. Each bin in the histogram corresponds to the coverage of one community. The bars with the same color correspond to the same contributing node. In Figure 4 (a) and (b), we show the subgraph before and after a critical tracing step. After that tracing step, the subgraph propagates to the upper part of the HoN. We can see that many nodes in the 8-th community are covered after this step, as indicated by the red arrow in Figure 4 (b). The node corresponding to the blue bars contributes most to the coverage of that community, as the blue bar in the 8-th bin is the tallest. By clicking on that blue bar, the contributing node is highlighted in yellow and the nodes reached from it are highlighted in blue in the subgraph. This indicates that the contributing node is an important transit point for the ships to propagate into the 8-th community. By identifying such nodes, domain experts can devise targeted species control strategies at certain critical ports to maximize the effectiveness and minimize the cost.

### 5.3 Aggregation View

The aggregation view provides an overview of the higher-order dependencies among groups of ports and their connections, corresponding to the design requirement **R2**. It also serves as a convenient interface to select the higher-order nodes with desired properties, e.g., the fifth-order nodes that contain ports in different eco-realms. The aggregation can be performed on the entire HoN or synchronized with the subgraph under expansion based on port grouping. The aggregated node corresponding to an original higher-order node is determined by converting each port associated with the higher-order node to the group containing that port. Formally, denoting a $k$-th-order node as a sequence of ports $\mathbf{n_i} = [p_{i_0}|p_{i_1}, \ldots, p_{i_{k-1}}]$, where $p_{i_0}$ is the current port and $p_{i_1}, \ldots, p_{i_{k-1}}$ are the previously visited ports, and the group of a port $p$ as $G(p)$, the aggregated node corresponds to node $\mathbf{n_i}$ can be written as

$$A(\mathbf{n_i}) = [G(p_{i_0})|G(p_{i_1}), \ldots, G(p_{i_{k-1}})]. \quad (3)$$

The edges are aggregated accordingly by summing up the weights of edges corresponding to the same pair of aggregated nodes.

We group the ports according to their eco-realms. This means that the higher-order nodes containing sequences of ports are aggregated into the higher-order nodes containing sequences of eco-realms. The edges are aggregated to show the number of ships moving among the eco-realms. Twelve groups of ports (i.e., eleven marine eco-realms and one group containing all freshwater ports) are considered. Unlike the original nodes, where two consecutive ports are always distinct, an aggregated node may contain two consecutive appearances of the same eco-realm, meaning that the ships move from one port to another in the same eco-realm. This will be effective for domain experts to distinguish the higher-order dependencies inside each eco-realm and among the eco-realms, which is critically important to the study of species invasion.

**Coarse Grouping Aggregation.** In some cases, the aggregation technique with the above *exact grouping* may not be necessary. For example, users may be interested in the higher-order nodes whose previous steps contain ports in other eco-realms without caring exactly what those eco-realms are. In other words, it suffices to distinguish the ports in the same eco-realm as the current port and the ports in different eco-realms. To accommodate this need, we further design an aggregation scheme with *coarse grouping*. With coarse grouping, the aggregated nodes can still be generated using Equation (3) but with a slightly different grouping function $G(p)$. Unlike the exact grouping function that always maps a port to a group, the coarse grouping function either maps a port to the group representing the eco-realm of the current port, or to a special status indicating that the port is in a different eco-realm. For example, the node [Singapore|Port Klang, Shanghai] will be aggregated into [Central Indo-Pacific|Central Indo-Pacific, Temperate Northern Pacific] with exact grouping but [Central Indo-Pacific|Central

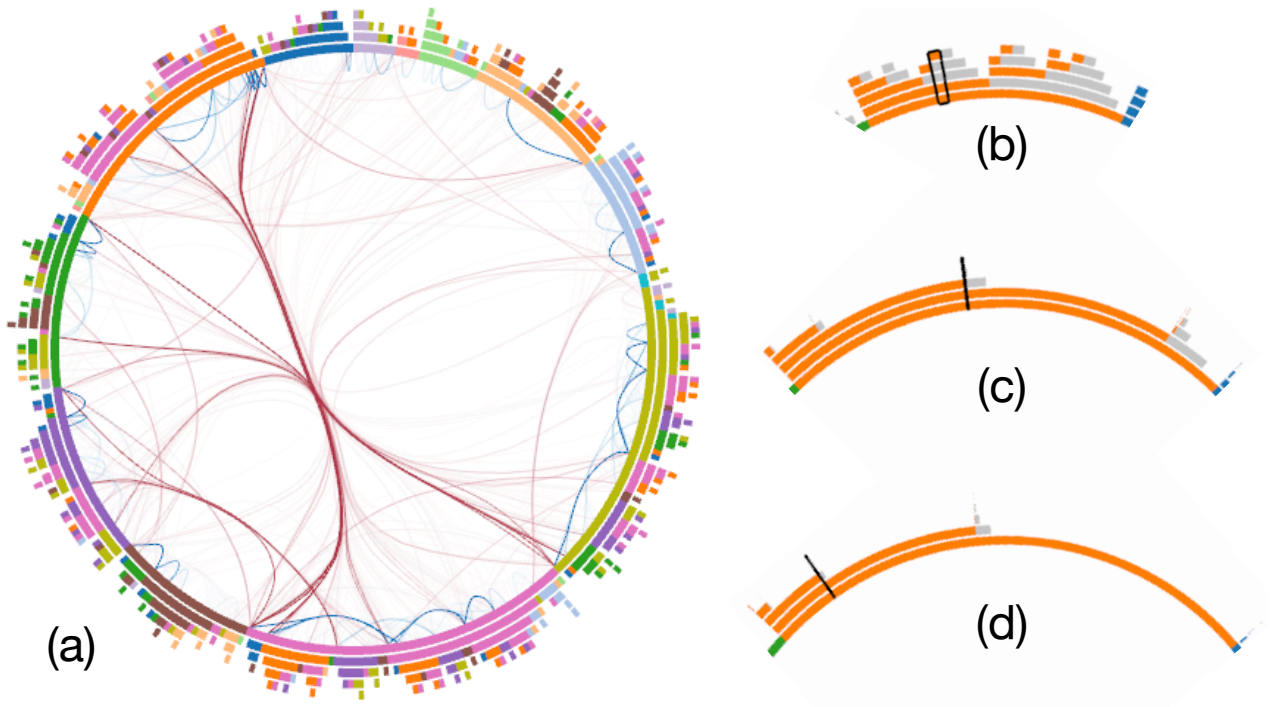

Figure 5: The aggregation view. (a) Exact grouping using eco-realms. (b) to (d) The eco-realm of "Temperate Northern Pacific" with coarse grouping. (b) Uniform node weight. (c) Nodes are weighted by the number of original nodes. (d) Nodes are weighted by the number of ships. The same aggregated node is highlighted in black in (b) to (d).

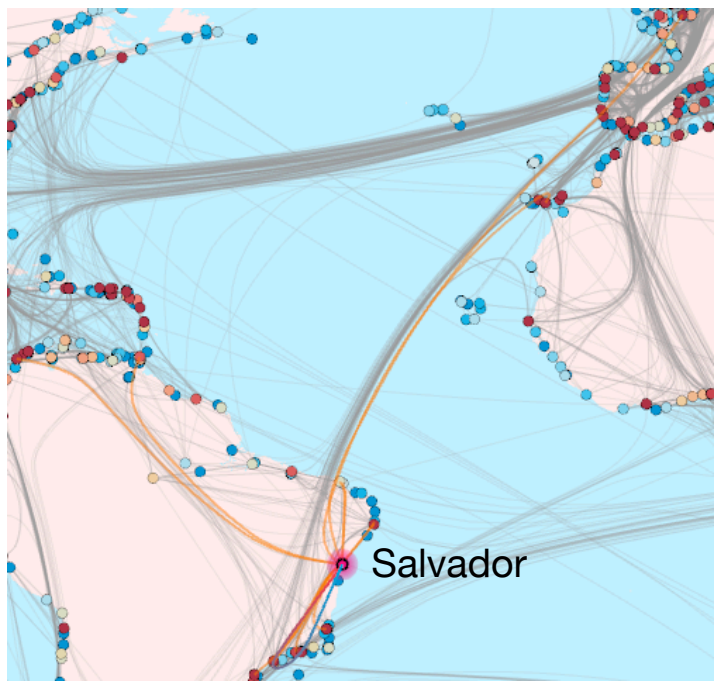

| Ports | #HO Nodes |
|---|---|
| Suape | 19 |
| Vitoria | 19 |
| Salvador | 13 |
| Tubarao | 6 |
| Praia Mole | 5 |
| Portocel | 2 |
| Ponta do Ubu | 2 |
| Aratu | 2 |
| Recife | 2 |
| Madre de Deus | 1 |
| Cabedelo | 1 |
| Ilheus | 1 |
| Maceio | 1 |
| Jubarte Field | 1 |

(a) (b)

Figure 6: Identifying a port of interest. (a) The port Salvador in Brazil is highlighted with a magenta halo in the geographic view. (b) The nearby ports are listed in the table view ordered by their numbers of associated higher-order nodes.

Indo-Pacific, Different Eco-realm] with coarse grouping. In our experiment, the number of aggregated nodes reduces from 396 to 180 with coarse grouping, allowing users to focus more on the between-group dependencies. Users can switch between exact grouping and coarse grouping depending on their needs.

**Network Layout.** We show the aggregation view using the circular layout, where the nodes are aligned on a circle and their connections are displayed inside the circle. The edges among nodes belonging to the same current group are colored in blue, while the edges among nodes belonging to different current groups are colored in brown. We bundle the edges for visual clarity using the force-directed edge bundling algorithm [13]. An aggregated node covers a sector of the circle, as highlighted by the black rectangle in Figure 5 (b). The number of layers in the sector represents the node's order, and the color of a layer represents the group of ports (i.e., eco-realm). The groups of ports are visited in the order from the outermost layer to the innermost layer (i.e., the current group is in the innermost layer). The gray color is reserved for the special group "different eco-realm". For example, the aggregated node highlighted in Figure 5 (b) exhibits five layers, from outermost to innermost, colored in orange, gray, gray, orange, and orange, respectively. This indicates that the node is fifth-order and the ships visited different eco-realms two and three steps before. The nodes are ordered according to their corresponding sequences of groups. That is, the nodes belonging to the same current group occupy a consecutive sector at the innermost layer, and then the nodes belonging to the same previous group are organized consecutively at the second inner layer, and so on. In Figure 5 (b), we can see that the nodes corresponding to the orange group are placed together. The second inner layer shows orange on the left side and gray on the right side, indicating that the nodes with the same previous group are on the left side and the nodes with different previous groups are on the right side.

The arc length of the sector is decided by the weight of the corresponding node. We provide three types of node weights. Figure 5 (b) shows the orange group with the uniform weight, where each node occupies the same arc length so that different nodes can be distinguished more easily. In Figure 5 (c), the aggregated nodes are weighted by the number of original nodes contained in them. We can observe from the arc lengths that most higher-order nodes exist among ports in the same eco-realm. In Figure 5 (d), the aggregated nodes are weighted by the number of ships related to each node. We observe that about half of the sector shows higher-order dependencies, within which a large proportion of ships travel within the same eco-realm, while a small proportion may bring in invasive species from other eco-realms, suggesting targeted control opportunities. A complete picture of the aggregation view with coarse grouping can be found in Figure 3 (d). Figure 5 (a) shows the aggregation view with exact grouping. Although it provides more details, it is more difficult to interpret as an overview due to its complexity.

## 6 Case Study on Species Invasion via Global Shipping Network

We worked in person with two domain experts in network science and marine ecology, and in this section we record the experts' workflow and observations when they first used HoNVis to explore the global shipping network. We then show how HoNVis reveals novel patterns at the global scale, which are valuable for decision-makers to devise effective species control strategies.

### 6.1 Data

Diverse types of data were used for this case study. The global ship movement data are made available by the Lloyd's List Intelligence, which contains more than two thirds of active ships globally (measured in dead weight tonnages). The raw data contain 3,415,577 individual ship voyages corresponding to 65,591 ships that move among 4,108 ports globally between May 1, 2012 and April 30, 2013. The data also contain metadata of ships, such as ship type, voyage start and end time, ship size, as well as metadata of ports such as coordinates and country. The environmental conditions (temperature and salinity) of ports are obtained from the Global Ports Database [17] and the World Ocean Atlas [4, 19]. The eco-region information comes from Marine Ecoregions of the World [25] and Freshwater Ecoregions of the World [1]. Ports (and associated ship movements) that have corresponding coordinates, eco-region and environmental conditions are retained for analysis.

### 6.2 Domain Experts' Workflow and Insights

**Locating Ports with Higher-Order Dependencies (R1).** The experts wanted to investigate potential species invasions from South America to Europe via global shipping, and evaluate the influence of higher-order movement patterns of shipping. As shown in Figure 6 (a), the experts first used the geographic view to zoom in to South America. To identify ports through which ships demonstrate higher-order movement patterns, the experts chose to color the ports by the number of higher-order dependencies, and focused on ports shown in red (the ones that demonstrate the most higher-order dependencies). The number of candidate ports is thus reduced from hundreds to tens. The experts then simply clicked on the area of interest, and in the table view (Figure 6 (b)), the ports in the area were sorted by the number of higher-order dependencies. The experts clicked through the top ports to highlight shipping paths from those ports, and quickly identified Salvador in Brazil, which shows a direct connection in the bundle from South America to Europe.

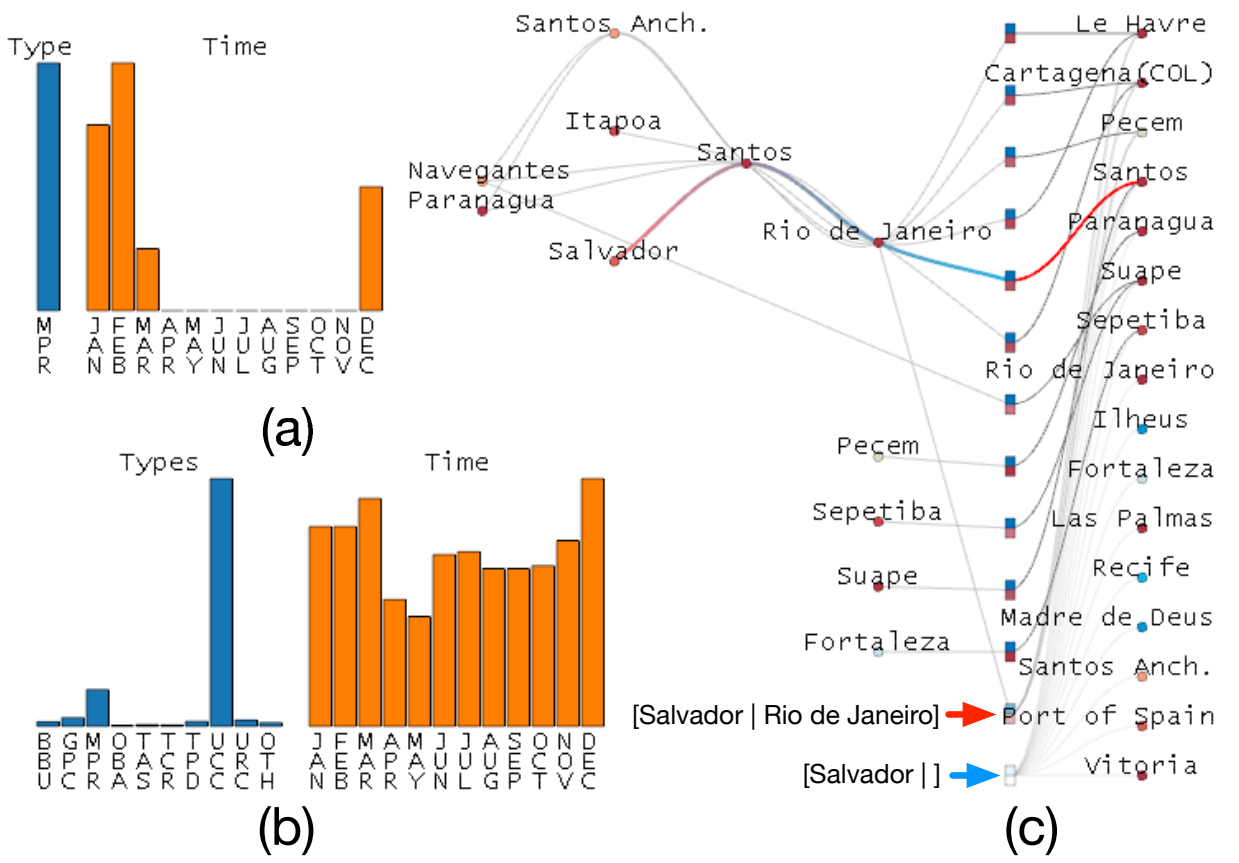

Figure 7: The higher-order dependencies related to Salvador. (a) Histograms of ship types and temporal activities of fourth-order movement patterns from Salvador. (b) Histograms of ship types and temporal activities for all ships from Salvador. (c) Higher-order dependencies related to Salvador in the dependency view.

**Exploring Higher-Order Dependencies (R3).** The experts then evaluated how the movement pattern from Salvador is influenced by from where the ships came to Salvador. After selecting Salvador in the table view, all its higher-order dependencies are displayed in the dependency view (Figure 7 (c)). At a glance, the experts knew that without knowing a ship's previous locations, the ship's next step from Salvador is uncertain. This is revealed by both the weak connections (dimmed visually) from the first-order node [Salvador|] (highlighted by the blue arrow in Figure 7 (c)) to all 16 potential destination ports on the right, and the white entropy box of [Salvador|] (high entropy indicating low certainty). A quick drag-and-drop selection of the destination ports reveals that the ships from Salvador are mainly container carriers (UCC), and shipping at Salvador remains active throughout the year (Figure 7 (b)).

Following the link from Rio de Janeiro to the second-order node [Salvador|Rio de Janeiro] (highlighted by the red arrow in Figure 7 (c)), the experts discovered that knowing ships came from Rio de Janeiro to Salvador does not significantly influence the ships' choices for the next step, indicated by the light red KLD box (meaning low difference compared with the distribution from the first-order node), and the light blue entropy box (indicating low certainty). Essentially, this implies that *the second order is insufficient in capturing the complex dependencies in this case*. It is likely that Rio de Janeiro, being the second largest city of Brazil, has a port so versatile and provides limited information in narrowing down complex ship movement patterns. The reason that the second-order node [Salvador|Rio de Janeiro] is included in HoN is that it bridges connections from other essential higher-order nodes.

The experts then proceeded to explore dependencies beyond the second order. By selecting the fourth-order path Salvador $\rightarrow$ Santos $\rightarrow$ Rio de Janeiro $\rightarrow$ Salvador, as highlighted in Figure 7 (c), the experts observed a *loop*, that if a ship has been observed following the loop at least once, the ship will keep following the loop for sure. The dark blue entropy box and dark red KLD box at port [Salvador|Rio de Janeiro, Santos, Salvador] indicate that this pattern displays high certainty and is significantly different than the first-order movement pattern. Moreover, the bar charts (Figure 7 (a)) in the dependency view show that ships following this fourth-order pattern are exclusively cruise ships (MPR) and are only active in the summer (December to March in the South Hemisphere), *revealing the underlying reason behind this higher-order dependency*.

**Exploring the Influence of Higher-Order Dependencies in Propagation (R4).** The experts further explored how higher-order dependencies influence the propagation of invasive species via shipping. Specifically, knowing that the ships came from Itapoa or Navengates before sailing through Santos and Rio de Janeiro to Salvador, the experts wanted to figure out how the species propagate differently. The experts first selected the fourth-order pathway Itapoa $\rightarrow$ Santos $\rightarrow$ Rio de Janeiro $\rightarrow$ Salvador in the dependency view, and the corresponding node [Salvador|Rio de Janeiro, Santos, Itapoa] is automatically selected in the subgraph view. The experts clicked "Trace" button to see how the species may propagate from the given port in a stepwise manner. As shown in Figure 8 (a), the species first went to Pecem in Brazil and then to New York City in USA. After that, with high certainty, the species were propagating toward the blue cluster on the left, which mainly consists of ports in Northeast America. After tracing a few more steps, the possible diffusion diverged. A branch kept propagating in Northeast America with high certainty. More interestingly, the species may influence multiple ports in East Asia, represented as the green cluster at the top-right corner as shown in Figure 8 (b), which was topologically far from the initial port Salvador on the lower left. The experts noticed the new spike in the stacked histogram, consisting mainly of a single color (blue). This indicates that a port is making significant contribution to the massive dispersion of species in that cluster. The experts clicked on the dominating blue bar of that spike, and the subgraph view reveals that Guangzhou was the port that facilitated the potential massive spread of invasive species in East Asia. *Knowing that Guangzhou is the entry point to species spreading in that region is vital when developing targeted invasive species control strategies* to prevent Brazilian species from invading East Asia. Tracing back, Guangzhou was invaded by ships sailing from Gibraltar through the Mediterranean Sea, then through the Suez Canal to the Red Sea, passing Jeddah and finally to Guangzhou. These ports on the shipping path also deserve close monitoring.

On the contrary, when the experts selected the pathway Navengantes $\rightarrow$ Santos Arch $\rightarrow$ Santos $\rightarrow$ Rio de Janeiro $\rightarrow$ Salvador in the dependency view, with high certainty the species will propagate toward the gray cluster at the bottom as shown in Figure 8 (c), which mainly consists of ports in Northwest Europe. The port leading to the mass diffusion in the cluster was Brunsbuttel. Through the interactive exploration and comparison, the experts gained a comprehensive understanding on how the higher-order dependencies may influence the subsequent propagation.

**Exploring Higher-Order Dependencies at Different Granularities (R2).** Finally, the experts wanted to explore the connections at a higher level: the *eco-realm* is the largest biogeographic division of the sea [25]; species coming from other eco-realms are more likely to be non-indigenous and will incur invasions. The question is: how do the connections differ whether the previous port was also in the Tropical Atlantic eco-realm (which Salvador is in) or was in a different eco-realm? The experts first chose to color the ports in the geographic view with eco-realms. Tropical Atlantic was colored dark green. Then the experts shifted to the aggregation view, and chose the sector which both the current and previous ports are in the Tropical Atlantic eco-realm. The sector is denoted by two layers of dark green, as shown in Figure 9 (a). The aggregation view reveals ampler and stronger intra-eco-realm connections as denoted by blue links, compared with inter-eco-realm connections as denoted by brown links (mainly connections to Temperate Southern Africa, Temperate Northern Atlantic, and Temperate South America). By cross-checking with ship types in the dependency view, the experts found out that variable types of ships exist for this case.

On the contrary, when the experts chose the sector which the current ports are in Tropical Atlantic but the previous ports are not (the sector denoted by the innermost layer as dark green and the outer layer as gray, as shown in Figure 9 (b)), the inter-eco-realm connections are stronger, including additional connections to Tropical Eastern Pacific and Central Indo-Pacific. Meanwhile, the dependency view suggested that these inter-eco-realm navigation patterns were exclusively made by container carriers. The experts came to the preliminary conclusion that *ships coming from different eco-realms were more likely to keep traveling among eco-realms, posing higher risks of bringing in non-indigenous species*. Furthermore, in terms of species management strategies for specific types of ships, *container carriers posed the highest risk for the introduction of non-indigenous species*.

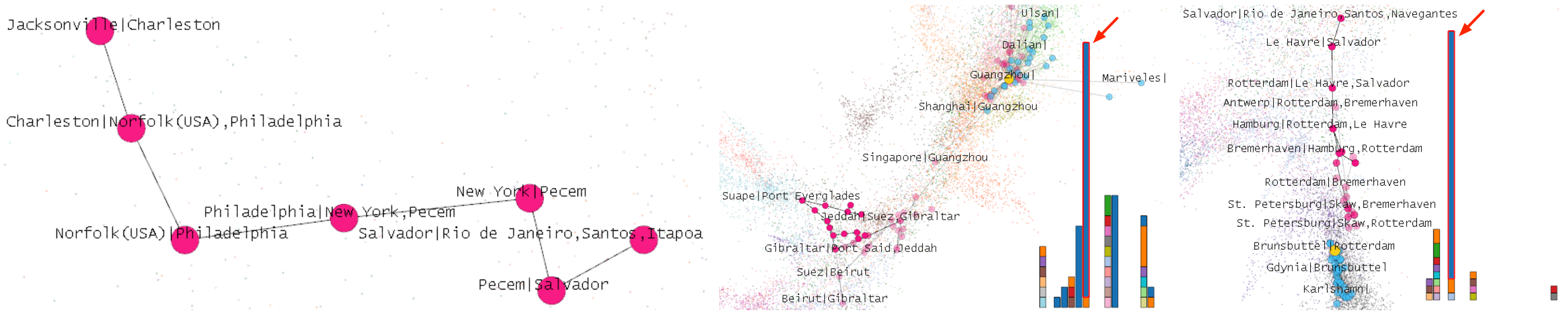

(a) (b) (c)

Figure 8: (a) Tracing how the species may propagate from Salvador in a stepwise manner. (b) The propagation eventually influences multiple ports in East Asia, which are far away from Salvador. (c) Another direction of the propagation covers multiple ports in Northwest Europe.

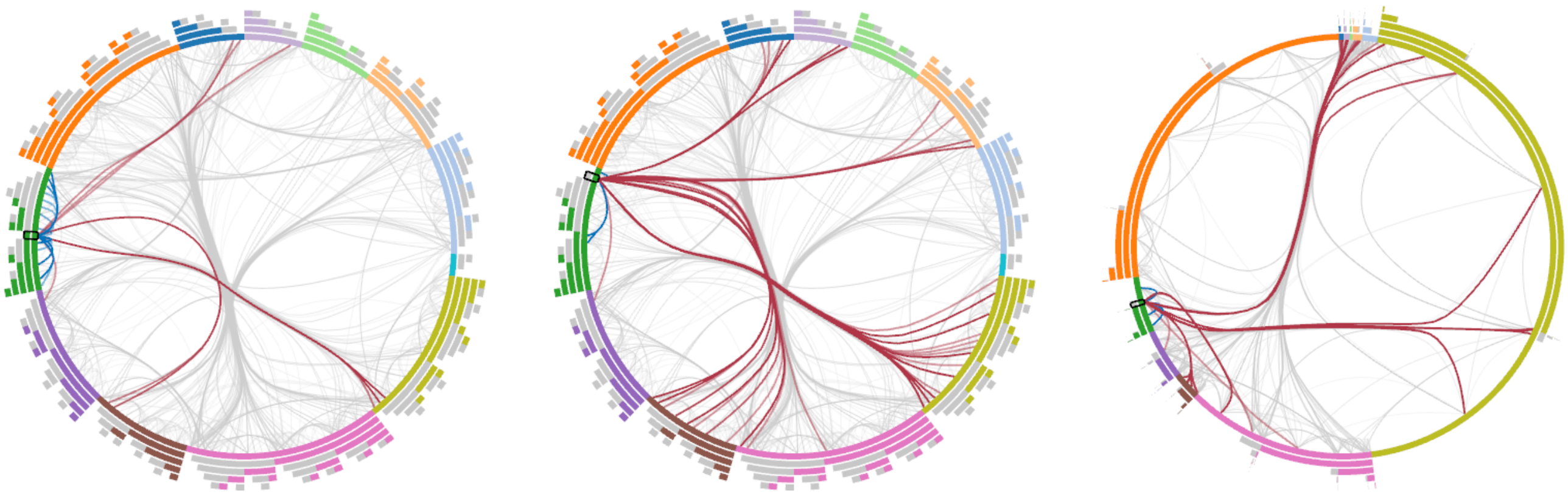

(a) (b) (c)

Figure 9: Investigating higher-order dependencies at different granularities. (a) Studying a sector which both the current and previous ports are in the Tropical Atlantic eco-realm. (b) Studying a sector which the current ports are in the Tropical Atlantic eco-realm, but the previous ports are not. (c) Changing the view in (b) from uniform node weight to weighted by the number of ships.

Last, the experts changed the widths of sectors from uniform to the number of ships, as shown in Figure 9 (c), which gives an intuitive overview of the composition of all higher-order dependencies. The experts noticed that although ships coming from other eco-realms to Tropical Atlantic have higher chances of keeping with the inter-eco-realm voyages, the number of inter-eco-realm trips was much less than that of intra-eco-realm trips. The fact that the more risky inter-eco-realm voyages were the minorities suggested that *targeted species control policies only need to focus on a small fraction of ships and routes*.

**Insights Revealed by HoNVis at the Global Scale.** HoNVis not only enables interactive exploration as shown in the above use case, but also reveals the influence of higher-order dependencies at the global scale. For example, one observation was for ports in the Arctic. The change of climate had been melting the Arctic sea ice at an alarming speed and opening up Arctic shipping routes [11]. Therefore, there are growing concerns on threats to the valuable resources in the Arctic posed by invasive species via the unprecedented growth of shipping. The PageRank algorithm naturally simulates the flow of species hitchhiking onto ships, with random resets accounting for the changing or unobserved shipping activities. The PageRank score of each port indicates the relative risk that species will end up to the port in multiple steps. The PageRank risk estimation on the FoN marks multiple ports in the Arctic as high risk, but as pointed out in Section 3, the HoN can improve the result of PageRank running upon. Surprisingly, the estimated risks for Arctic ports were overwhelmingly overestimated on the FoN. This is indicated by the ports in blue as shown in Figure 10. For example, the PageRank score of Murmansk, a major Arctic port in Russia, was $4.52 \times 10^{-4}$ on the FoN, but only $1.57 \times 10^{-4}$ on the HoN. The dependency view suggested that by using the HoN, traffic from hub ports such as Rotterdam to the Arctic ports is more likely to go back immediately to those hub ports rather than moving randomly among Arctic ports. Thus the relative flow of species in the Arctic is smaller on HoN. The information on the overestimation of risks made possible by HoNVis is important for policy makers.

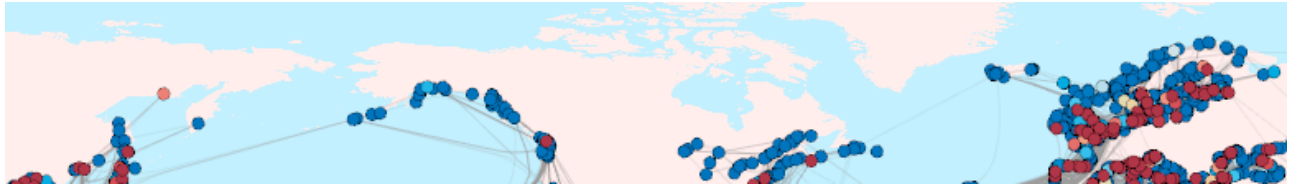

Figure 10: Comparison of PageRank risk simulation on the FoN and the HoN. Blue ports are risks overestimated on the FoN and red ports are risks underestimated on the FoN.

## 7 Conclusions and Future Work

We have presented HoNVis, a visual analytics framework for visualizing and exploring higher-order networks. We focus on the global shipping network and work closely with domain experts in network science and marine ecology to compile the task list and define design requirements. Our HoNVis design leverages five linked views to enable users to explore the HoN at different levels of de-tail and investigate higher-order dependencies among higher-order nodes. By directly contrasting the HoN and its FoN counterpart and visualizing higher-order dependencies, we tackle the key challenges in visualizing higher-order dependencies in networks, which is a milestone in pushing the understanding of the formation and impact of higher-order dependencies. The efficacy of HoNVis is demonstrated through results gathered by two domain experts who use the system to investigate species invasion in the global shipping network. Several critical insights that can only be obtained with the use of HoNVis are reported.

We acknowledge the limitations of the current version of HoNVis, including the lack of effective visual hints to aid the users in navigating through the different views, and the challenge of labeling when the data are large. We advocate the idea of automatically producing statistics of all possible dependency structures (such as large loops) and aiding in the identification of principal patterns, which is a non-trivial task given the computational complexity.

Besides the application in global shipping and species invasion, the framework of HoNVis can be generalizable to other types of HoNs, which we plan to implement in the near future. For example, given that air transportation exhibits higher-order dependencies [24], HoNVis can help to explore epidemic outbreak scenarios through domestic and international travels, by substituting ships with airplanes and invasive species with contagious diseases. Similarly, HoNVis can also help to explore information diffusion patterns through phone call or online activities in social networks, by treating phone call or retweet cascades as ship trajectories.

## Acknowledgements

This work was supported in part by the U.S. National Science Foundation through grants IIS-1456763, IIS-1455886, ACI-1029584, DEB-1427157, and the U.S. Army Research Laboratory under Cooperative Agreement No. W911NF-09-2-0053.